\documentclass[epj]{svjour}
\usepackage{graphicx}	% Include figure files
\usepackage{url}
\usepackage{hyperref}
\usepackage[utf8]{inputenc}
\usepackage{xspace}	% Include xspace
\usepackage{amsmath}	% Include extra math stuff
\usepackage{comment}
\begin{document}
\title{New, multipole solutions of relativistic, viscous hydrodynamics}
\author{Tamás Csörgő\inst{1,2} \and Gábor Kasza\inst{1,2,3}% etc
}                     % Do not remove
\offprints{}          % Insert a name or remove this line
\institute{Wigner Research Centre of Physics, H-1525 Budapest 114, P.O.Box 49, Budapest, Hungary \and
Eszterh\'azy K\'aroly University, K\'aroly R\'obert Campus,
H-3200 Gy\"ongy\"os, M\'atrai \'ut 36, Hungary \and
E{\"o}tv{\"o}s Lor{\'a}nd University, H-1117 Budapest, P{\'a}zm{\'a}ny P.~s.~1/A, Hungary 
}
\date{Received: date / Revised version: date}
\abstract{
We present a new class of exact fireball solutions of relativistic dissipative hydrodynamics. We describe new exact solutions both
for the relativistic Navier-Stokes and for the Israel-Stewart theory, for arbitrary shear and bulk viscosities, as well as for other dissipative coefficients. The common property of these solutions is the presence of the relativistic Hubble flow.
Our results generalize the recently found first solution in these classes, for an arbitrary
temperature dependent speed of sound, shear and bulk viscosity, heat conduction and fluctuating initial temperature profiles.
These solutions are causal and  not only stable but also asymptotically perfect. A strong and narrow peak in the kinematic bulk viscosity
is shown to imitate the effects of a first order phase transition. The new class of asymptotically perfect solutions is thus found to be very rich, but at the same time mostly academic as the solutions are limited by the spherical symmetry of the Hubble flow field.
\PACS{
      {PACS-key}{discribing text of that key}   \and
      {PACS-key}{discribing text of that key}
     } % end of PACS codes
} %end of abstract
\maketitle
\section{Introduction}
\label{intro}
The Navier-Stokes equations govern  the flow of fluids such as water and air. Relativistic fluids, such as the nearly perfect liquid
created in the relativistic collisions of heavy nuclei like gold and lead at the Relativistic Heavy Ion Collider at BNL, USA
and at the Large Hadron Collider at CERN, respectively, are  best described with the help of various solutions of generalized
relativistic dissipative hydrodynamical equations.

However, there is no proof even for the most basic questions one can ask: do exact solutions of the Navier-Stokes equations exist, 
and if yes, are they unique? The proof for the existence and uniqueness of these solutions for arbitrary initial conditions
is one of the Millennium Problems of the Clay Mathematical Institute  (CMI)~\cite{CMI:Navier-Stokes}.
Such a proof could provide not only certitude, but also understanding:
although the Navier-Stokes equations were written down in the 19th Century, 
our understanding of them remains minimal so far. The situation is even worse in the domain of viscous hydrodynamics of relativistic flows,
where, in contrast to the non-relativistic domain,  there is not even a consensus on the exact form of the equations that one should solve
to understand relativistic dissipative hydrodynamics~\cite{deSouza:2015ena}.

This is the motivation why in this extended manuscript we provide exact solutions of two forms of dissipative relativistic hydrodynamics.
We solve not only the so called relativistic Navier-Stokes equations, but also provide solutions to another form of dissipative relativistic hydrodynamics, the so called Israel-Stewart theory.

We describe exact solutions of the relativistic NS equations, 
where there is a great amount of freedom in the choice of the average value, or, of the temperature dependence 
of the shear and bulk viscosity as well as the heat conductivity coefficients, and the initial density or temperature profile
may also have arbitrary fluctuations. However, in each case, we provide solutions  that include a special flow velocity profile, 
the so called Hubble flow. For expansions with only one (longitudinal) spatial dimensions, this Hubble flow field coincides with the so called
Hwa-Bjorken flow field of refs.~\cite{Hwa:1974gn,Bjorken:1982qr}, very well known in the field of ultra-relativistic heavy ion collisions.

In physical cosmology, in 1 + 3 dimensions, the Hubble flow profile has a renowned cosmological and astrophysical relevance,
as it corresponds to the velocity profile of the galaxies in our Universe after the Big Bang
~\cite{Friedman:1922kd,Hubble:1929ig,Livio:2013RAG}. Although it is not very well known, a Hubble flow also  describes reasonably well the final velocity profile of the Little Bangs of relativistic heavy ion collisions as clear from the comparison of the experimental data with solutions that used the Hubble profile~\cite{Csorgo:2003ry,Csanad:2004mm,Chojnacki:2004ec,Csorgo:2003rt,Csanad:2014dpa,Shi:2014kta,Csanad:2015xra}.
The flow field in all of these solutions goes back to the Hubble flow ${\mathbf v} = H(t) {\mathbf r}$.
These solutions were  naturally extended to describe ellipsoidal, directional Hubble flows with a constant, average value of the speed of sound in ref.~\cite{Akkelin:2000ex}. This solution was generalized for an arbitrary temperature dependent speed of sound 
in the case of non-vanishing, conserved baryon charges ($\mu_B \neq 0 $), see ref.~\cite{Csorgo:2001xm}. 
However, at present the lattice QCD equation of state is known precisely only at valishing net baryon density~\cite{Borsanyi:2010cj}. 
Hubble flows with temperature dependent speed of sound, consistent with the lattice QCD Equation of State
at a vanishing net baryochemical potential were found
in the relativistic kinematic domain in ref.~\cite{Csanad:2012hr}.
From this paper on, the temperature dependence of the speed of sound was handled similarly in both 
the baryon-free region and in the case of non-vanishing
conserved (baryon)charge, see  for example in the new rotating and directionally Hubble - expanding solutions of ref.~\cite{Csorgo:2013ksa}.

An exact solution of the relativistic Navier-Stokes equations was reported in ref.~\cite{Csanad:2019lcl}
for an arbitrary shear viscosity and heat conductivity, using several different classes for the bulk viscosity,
for a temperature independent (constant) speed of sound and initially homogeneous temperature profile.
Here we present the  generalization of  the solutions of ref.~\cite{Csanad:2019lcl}, to an arbitrary
temperature dependent speed of sound, shear and bulk viscosity, heat conduction as well as a fluctuating initial temperature and
correspondingly fluctuating density profiles. Given that  our solutions too are valid for  the Hubble flow profile, 
they are apparently not only causal, but also stable for small density and corresponding
temperature perturbations. 
Furthermore, these solutions are  also asymptotically perfect in the sense that their time evolution approaches the Csörgő - Csernai - Hama - Kodama (CCHK) perfect fluid solution~\cite{Csorgo:2003ry}, but starting from a lower initial temperature. During the time evolution, the temperature each viscous case is gradually heated up to reach the time-evolution
of the asymptotically equivalent CCHK perfect fluid solution, as we show below.

In this extended manuscript we also describe the solutions of the so-called Israel-Stewart (IS) theory of dissipative relativistic hydrodynamics, and demonstrate that these solutions are also in the class of asymptotically perfect solutions. This is because some solutions of the relativistic NS
theory have problems with violations of causality and stability, as reviewed recently in ref.~\cite{deSouza:2015ena}. Surprisingly we find that the
new exact solutions of the IS theory that we describe in this extended manuscript have several properties common with the exact solutions of the NS
theory of relativistic dissipative hydrodynamics. Namely,
these new solutions of the IS theory are  also asymptotically perfect: their time evolution approaches the Csörgő - Csernai - Hama - Kodama (CCHK) perfect fluid solution of ref.~\cite{Csorgo:2003ry}. 

Thus the class of asymptotically perfect solutions of relativistic dissipative hydrodynamics is thus  apparently and interestingly rather broad.
This is why we consider that our central result is the discovery of the class of asymptotically perfect solutions of dissipative relativistic hydrodynamics. This class, as we shall show below, includes solutions with fluctuating initial temperature and corresponding (entropy)density profiles, temperature dependent shear and bulk viscosity coefficients, temperature dependent speed of sound and solutions of both the Navier-Stokes and the Israel-Stuart theory of dissipative relativistic hydrodynamics. The main limitation of the presented solutions seems to be the spherical symmetry of the Hubble flow field, which is a common part of all of the solutions described in this work. It is a research topic to find new exact
and asymptotically perfect solutions of the theories of dissipative relativistic hydrodynamics that may violate the spherical symmetry of the Hubble flow field.

The relativistic Navier-Stokes (NS) equations are
one of the possible theoretical formulations of dissipative relativistic  hydrodynamics.
One of their advantage is they have a clear non-relativistic limit:
the Navier-Stokes equations~\cite{deSouza:2015ena}. However,  it is also known that this theory
allows for acausal propagation with speed that is larger than the speed of light and also this theory
has certain unstable solutions and modes, as reviewed recently in ref.~\cite{deSouza:2015ena}.

%This property is inherited by the solutions that we present in this work: in the non-relativistic kinematic domain,
%these relativistic solutions provide an exact solution of the non-relativistic Navier-Stokes equations. We emphasize throughout this manuscript that
%the new class of solutions is a class of asymptotically perfect solutions, and they at late times they all tend to the exact solutions of relativistic perfect fluid hydrodynamics, the so called CCHK solution described in ref. ~\cite{Csorgo:2003ry}. This perfect fluid solution in the non-relativistic kinematic limit also tends to an exact solution of non-relativistic fireball hydrodynamics~\cite{Csorgo:2001ru}, as was shown in ref.~\cite{Csorgo:2003ry}.

Let us also note that we will describe also new exact solutions of the relativistic Israel-Stewart hydrodynamics, that is causal and in certain domains of the parameter space, also stable~\cite{deSouza:2015ena}. However,   the newly found exact solutions of the  IS theory, described in this manusript,  can not correspond to the exact solutions of the non-relativistic Navier-Stokes equations in their non-relativistic limit, as they also depend on an additional relaxation time parameter $\tau_\Pi$, that is not part of the original
non-relativistic Navier-Stokes theory. 
More detailed investigation of the non-relativistic limit of our new solutions however is well beyond the scope of the current manuscript.

We emphasize, that in this work we provide exact solutions of the equations of relativistic dissipative hydrodynamics, which is a theoretical result of certain academic interest. We however do not yet compare our results with experimental data, as this would go very much beyond the scope of the present manuscript. So the modelling aspect of the results is not emphasized in this manuscript. Instead, we emphasize our desire to know and understand as well as to deepen our level of understanding of a very deep theoretical problem, not only in 1+1 but also in 1+ 3 or in general 1+ d dimensions. 

\section{Relativistic hydrodynamics}
\label{sec:2}
The dynamical equations of motion of relativistic fluids correspond to the local conservation of four momentum and the continuity equation of the conserved density $n$:
\begin{align}
    \partial_{\mu}\left(nu^{\mu}\right) &= 0,\\
    \partial_{\mu}T^{\mu \nu} &= 0,
\end{align}
where $T^{\mu\nu}$ is the energy-momentum tensor:
\begin{equation}\label{eq:energy-mom_tensor}
    T^{\mu \nu} = \left( \varepsilon + p \right) u^{\mu}u^{\nu} - p g^{\mu \nu} + q^{\mu} u^{\nu} + q^{\nu} u^{\mu}+\pi^{\mu \nu}.
\end{equation}
The metric tensor of Minkowski space is denoted by $g^{\mu\nu}=\textnormal{diag}\left(1,-1,-1,-1\right)$, the four-velocity by $u^{\mu}$, the pressure by $p\equiv p(x)$, the energy density by  $\varepsilon\equiv \varepsilon(x)$ and the temperature field by $T\equiv T(x)$. These fields are functions of the $x\equiv x^{\mu}$ four-coordinate. The first two terms of eq.~\eqref{eq:energy-mom_tensor} describe perfect fluids, while the additional terms stand for the dissipative, viscous effects. The heat current $q^{\mu}$ and the heat conductivity $\lambda$ are related:
\begin{equation}\label{eq:q_mu}
    q^{\mu}=\lambda\left(g^{\mu \nu}-u^{\mu}u^{\nu}\right)\left(\partial_{\nu}T-T u^{\rho}\partial_{\rho}u_{\nu}\right),
\end{equation}
The relativistic Navier-Stokes form of %the stress tensor 
$\pi^{\mu\nu}$ reads as:
\begin{equation}\label{eq:pi_munu}
\begin{split}
    \pi^{\mu \nu}=&\eta \left[\left(g^{\mu\rho}-u^{\mu}u^{\rho}\right)\partial_{\rho}u^{\nu} + \left(g^{\nu\rho}-u^{\nu}u^{\rho}\right)\partial_{\rho}u^{\mu} \right]+\\ &\left(\zeta-\frac{2}{d}\eta\right)\left(g^{\mu\nu}-u^{\mu}u^{\nu}\right)\partial_{\rho}u^{\rho}.
\end{split}
\end{equation}
The shear and bulk viscosities are denoted by $\eta$ and $\zeta$ while the number of spatial dimensions by $d (= 3)$. 
For us, a particularly interesting advantage of this  form of dissipative relativistic hydrodynamics is that
it apparently reproduces the Navier-Stokes equations in the non-relativistic $|\mathbf v| \ll 1$, $\gamma= 1/\sqrt{1 - {\mathbf v}^2} \rightarrow 1$ limit. The detailed discussion of the conjectured non-relativistic limit of our relativistic solutions 
goes however beyond the scope of the present manuscript.
In the case of perfect fluids, the non-relativistic solution of the non-relativistic limit of the relativistic CCHK solution was discussed in 
Section 7 of ref.~\cite{Csorgo:2003ry}.

At this point let us note, that the relativistic Navier-Stokes equations may lead to certain solutions, that violate causality~\cite{deSouza:2015ena}. 
However, causality is violated by some well-known solutions of Einstein's equations of general relativity as well, 
see for example G\"odel's famous solution~\cite{Godel:1949ga}, which nevertheless
did not call into question Einstein's general theory of relativity, but rather helped to understand its properties better.
By this analogy, we argue that it is important to investigate not only the properties of the relativistic Navier-Stokes equation
but it is even more important to scrutinize the properties of a given solution. The particular solutions that we discuss in this manuscript all share the Hubble form of the four-velocity field, which is well known to be a causal solution.

The above set of partial differential equations of relativistic hydrodynamics is closed by the equation(s) of state (EoS).
In this work we assume, following ref.~\cite{Csorgo:2001xm,Csanad:2012hr,Csorgo:2016ypf,Nagy:2016uiz,Csorgo:2018tsu}, 
that the EoS is 
\begin{equation}
    \varepsilon = \kappa(T) p.
\end{equation}
For a baryon-free evolution, $\mu= 0$, and in this case the coefficient $\kappa(T)$
is related to the temperature dependence of speed of sound as $\kappa(T)=c_s^{-2}(T)$. 
The simplest choice is the temperature independent, average speed of sound. For example, dust corresponds to $c_s^2=0$, the ultra-relativistic or non-relativistic ideal gas corresponds to the $\kappa = 3$ and $\kappa = 3/2$ cases, respectively. The same  EoS is applied by some renowned solution of relativistic hydrodynamics like refs.~\cite{Bjorken:1982qr,Belenkij:1956cd} or the cosmological solutions of the famous Friedmann equations for radiation and matter dominated Universes \cite{Friedman:1922kd}. 
Several well-known exact solutions of perfect fluid hydrodynamics utilize an average, temperature independent value of the speed of sound, see for example refs.~\cite{Bondorf:1978kz,Csorgo:2006ax,Csorgo:2003ry,Csorgo:2003rt}, while several other exact solutions utilize  a temperature
dependent speed of sound, corresponding to  a $\kappa(T)$ function, see for example refs.~\cite{Csorgo:2001xm,Csanad:2012hr,Csorgo:2016ypf,Nagy:2016uiz,Csorgo:2018tsu}.

\section{New viscous, Hubble-type solutions}
\label{sec:3}
We search such solutions where the velocity field is a relativistic Hubble flow: 
\begin{equation}\label{eq:hubble-flow}
    u^{\mu}=\frac{x^{\mu}}{\tau}=\gamma \left(1,\frac{r_x}{t},\frac{r_y}{t},\frac{r_z}{t}\right).
\end{equation}
where $\gamma=t/\tau$, $t=x^0$ and $\tau=\sqrt{t^2-r_x^2-r_y^2-r_z^2}$ is the proper time.
Let us search for self-similar exact solutions. %with this flow profile.
The scaling variable $s$   satisfies the 
$u^{\mu}\partial_{\mu} s = 0$ condition. For the Hubble flow, the scaling variable $s$ can be
any function of the directional scaling variables 
\begin{equation}
    s_x=\frac{r_x}{t},\:s_y=\frac{r_y}{t},\:s_z=\frac{r_z}{t},
\end{equation}
which satisfy the scale equation separately:
\begin{equation}
    u^{\mu}\partial_{\mu}s_i = \partial_{\tau} s_i=0, \quad\quad \mbox{\rm for} \quad i= (x, y, z),
\end{equation}
For a Hubble-flow profile, the great simplification is that in this case, the relativistic Navier-Stokes equations are reduced to the $p\tau-\zeta d=\phi(\tau)$ condition \cite{Csanad:2019lcl}, where $\phi$ is an arbitrary function of the proper time. 
The continuity equation of particle density is
\begin{equation}\label{eq:continuity_with_hubble}
    \partial_{\tau}n+\frac{d}{\tau}n=0.
\end{equation}

The energy conservation can be expressed in terms of pressure as:
\begin{equation}\label{eq:energy-cons_with_hubble}
    \partial_{\tau}p+\left(1 + \frac{1}{\kappa}\right) \frac{d}{\tau} p =\frac{d^2}{\tau^2}\frac{\zeta}{\kappa}.
\end{equation}
Notice, that the effects of shear viscosity and  heat conductivity n cancel, because
~\eqref{eq:continuity_with_hubble} 
and
~\eqref{eq:energy-cons_with_hubble} 
do not depend on $\lambda$ and $\eta$~\cite{Csanad:2019lcl}.

One can also derive the entropy equation by using the $p=\varepsilon/\kappa$ and the $d\varepsilon = T d\sigma$ equalities in eq.~\eqref{eq:energy-cons_with_hubble}:
\begin{equation}\label{eq:entropy_with_hubble}
    \partial_{\tau}\sigma+\frac{d}{\tau}\sigma = \frac{d^2}{\tau^2}\frac{\zeta}{T} \ge 0.
\end{equation}

For a non-vanishing net baryon charge scenario, we consider equations ~\eqref{eq:continuity_with_hubble} 
and
~\eqref{eq:energy-cons_with_hubble},
while for the case of a vanishing net baryon charge,
we solve equations~\eqref{eq:entropy_with_hubble},
and~\eqref{eq:energy-cons_with_hubble}.
Both cases 
correspond to a coupled set of ordinary differential equations.
In both cases, we have thus reduced the complex set of partial differential equations of relativistic hydrodynamics to
a system of first order, coupled, ordinary differential equations, 
that can be readily solved with generally accessible software packages like Maple or Mathematica.

In this work, we search for exact analytic solutions, and to facilitate that goal we assume, 
that the ratio of the bulk viscosity to pressure is a constant ($\zeta \propto p$):
\begin{equation}\label{eq:zeta}
    \zeta=\zeta_0 \frac{p}{p_0},
\end{equation}
where $\zeta_0$ and $p_0$ are the initial values of the bulk viscosity and the pressure.
In case of a vanishing baryonchemical potential, this assumption corresponds to a temperature dependent
ratio of $\zeta/\sigma$. 

In our solutions, the shear viscosity and heat conductivity effects  cancel exactly, similarly to the solutions of ref.~\cite{Csanad:2019lcl}.

The solution of eq.~\eqref{eq:energy-cons_with_hubble} for the pressure is
\begin{eqnarray}
    p(\tau)&=&p_0  \left(\frac{p_A}{p_0}\right)^{1 - \frac{\tau_0}{\tau}} \left(\frac{\tau_0}{\tau}\right)^{d\left(1+\frac{1}{\kappa}\right)}
                       \\
    \frac{p_A}{p_0}&=&f_{A,0} \, = \, \exp\left[\frac{d^2 \zeta_0}{\kappa_0 p_0 \tau_0}\right], \label{eq:pressure_scale}
\end{eqnarray}
where $\tau_0$ stands for the initial proper time, $\kappa_0$ is a constant and we introduced $f_{A,0} \ge 1$ as
the ratio $p_A$  to the initial pressure $p_0$. The initial pressure of an  asymptotic perfect fluid CCHK solution
is denoted by $p_A$.
For late times, $\tau \gg \tau_0$, this solution of the relativistic Navier-Stokes equation approaches the perfect CCKH solution of ref.~\cite{Csorgo:2003ry} with an initial pressure $p_A$.

\section{Multipole solutions with a conserved charge}
\label{sec:4}
If the bariochemical potential is finite ($\mu\neq 0)$ and the pressure is 
\begin{equation}
    p=nT,
\end{equation}
then eq.~\eqref{eq:continuity_with_hubble} can be solved~\cite{Csorgo:2003rt,Csanad:2012hr,Csanad:2004mm} as
\begin{equation}\label{eq:part_density}
    n=n_0 \left(\frac{\tau_0}{\tau}\right)^{d}\mathcal{V}\left(s_x,s_y,s_z\right),
\end{equation}
where $\mathcal{V}\left(s_x,s_y,s_z\right)$ is an arbitrary function of an arbitrary combination of $s_x$, $s_y$ and $s_z$. The equation for the pressure~\eqref{eq:energy-cons_with_hubble} can be rewritten as an equation for the temperature, and solved as
\begin{equation}\label{eq:temp_solution}
    T=T_0 \left(\frac{T_A}{T_0}\right)^{1-\frac{\tau_0}{\tau}}  \left(\frac{\tau_0}{\tau}\right)^{\frac{d}{\kappa_0}}\mathcal{T}\left(s_x,s_y,s_z\right),
\end{equation}
where the initial temperature denoted by $T_0$ and the asymptotic perfect fluid initial temperature by $T_A$,
\begin{equation}
    \frac{T_A}{T_0} = \frac{p_A}{p_0} \, = \, f_{0,A} \, = \, \exp\left(\frac{\zeta_0d^2}{\kappa_0 p_0 \tau_0}\right).
    \label{eq:temp-solution-zeta0}
\end{equation}
In eq.~(\ref{eq:temp_solution}), 
$\mathcal{T}$ is an arbitrary function of the scale variables, related to $\mathcal{V} $ with a matching condition as
\begin{equation}\label{eq:matching_cond}
    \mathcal{T}\left(s_x,s_y,s_z\right)=\frac{1}{\mathcal{V}\left(s_x,s_y,s_z\right)}.
\end{equation}

\section{Multipole solutions without conserved charge}
\label{sec:5}
When  $\mu=0$, there is no conserved charge,  and the pressure is obtained from thermodynamics as $\varepsilon+p=T\sigma$ :
\begin{equation}
    p=\frac{T\sigma}{1+\kappa}.
\end{equation}
Let we solve eq.~\eqref{eq:entropy_with_hubble} for the entropy density, together with the 
energy equation. For an average $\kappa=\kappa_0$
\begin{equation}\label{eq:asymp_sigma}
    \sigma=\sigma_0     \left(\frac{\sigma_A}{\sigma_0}\right)^{1-\frac{\tau_0}{\tau}} \left(\frac{\tau_0}{\tau}\right)^{d}
    \mathcal{V}\left(s_x,s_y,s_z\right).
\end{equation}
where $\sigma_A$ is the initial entropy density of the asymptotic perfect fluid solution in the $\tau\rightarrow \infty$:
\begin{equation}\label{eq:entropy_scale}
   \frac{\sigma_A}{\sigma_0}  \, = \, f_{0,A}^{\frac{\kappa_0}{1 + \kappa_0}} \, = \, \exp\left(\frac{\zeta_0d^2}{p_0\tau_0}\frac{1}{1+\kappa_0}\right).
\end{equation}
and $\mathcal{V}\left(s_x,s_y,s_z\right)$ satisfies the matching condition of eq.~\eqref{eq:matching_cond}. The temperature field has the same form as in eq.~\eqref{eq:temp_solution}, but with the following  expression of $T_A$:
\begin{equation}
    \frac{T_A}{T_0} =  \left(\frac{\sigma_A}{\sigma_0} \right)^{\frac{1}{\kappa_0}}  \, = \, 
        \exp\left(\frac{\zeta_0d^2}{\kappa_0 p_0\tau_0}\frac{1}{1+\kappa_0}\right).   
\end{equation}

\section{Temperature dependent speed of sound}
\label{sec:6}
At $\mu=0$, the entropy and the temperature equations, without our ansatz that says $\zeta \propto p$ read as:
\begin{align}
    &\partial_{\tau}\sigma+\frac{d}{\tau}\sigma = \frac{d^2}{\tau^2}\frac{\zeta}{T},\label{eq:entropy_equation_kT_mu_eq_0}
    \\
    &\frac{1+\kappa}{T}\left[\frac{d}{dT}\left(\frac{\kappa T}{1+\kappa}\right)\right]\partial_{\tau}T + \frac{d}{\tau}=\frac{d^2}{T\tau^2}
    \frac{\zeta}{\sigma} .\label{eq:energy_equation_kT_mu_eq_0}
\end{align}
This is the general form of the equations, for any temperature dependent speed of sound $c_s^2(T) \equiv \frac{1}{\kappa(T)}$
and for any temperature dependent
kimenatic bulk viscosity $\frac{\zeta}{\sigma} \equiv \frac{\zeta}{\sigma} (T)$.
Eqs.~(\ref{eq:entropy_equation_kT_mu_eq_0}-\ref{eq:energy_equation_kT_mu_eq_0}) can be further simplified in our  case of $\zeta = \zeta_0 \frac{p}{p_0}$ and $\kappa  \equiv \kappa(T)$ by using
\begin{equation}
\frac{1}{T} \frac{\zeta}{\sigma} \equiv
    \frac{1}{T} \frac{\zeta}{\sigma}\left(T\right) \, = \, 
    \frac{\zeta_0}{p_0}\frac{1}{1+\kappa(T)}. \label{eq:simplified-kinematic-bulk}
\end{equation}
For  a more realistic temperature dependent speed of sound,  lattice QCD simulations \cite{Borsanyi:2010cj} 
provide the temperature dependence of $\kappa(T) = \frac{\epsilon(T)}{p(T)}$.
%suggest a peak in the temperature dependence of $\kappa$ in the vicinity of the critical temperature of the QGP (Quark Gluon Plasma).
Although definitive lattice QCD results are not yet available   at $\mu \neq 0$,
the temperature equation can be rewritten in terms of $\kappa(T)$
in this case too, as follows:
\begin{equation}\label{eq:energy_equation_kT_mu_neq_0}
    \frac{1}{T}\left[\frac{d}{dT}\left(\kappa T \right)\right]\partial_{\tau}T + \frac{d}{\tau}=\frac{d^2}{\tau^2}\frac{\zeta_0}{p_0}.
\end{equation}

We thus have reduced the complicated partial differential equations of hydrodynamics to first order, ordinary differential equations for any  $\kappa(T)$ function. 
In this work, we do not detail further the exact but complicated solutions of eqs.~\eqref{eq:energy_equation_kT_mu_eq_0}-\eqref{eq:energy_equation_kT_mu_neq_0},
as these ordinary differential equations can be solved readily with the available numerical packages.
A possible choice for $\kappa(T)$ is a parametrization of the lattice QCD equation of state~\cite{Borsanyi:2010cj},
based on hydrodynamical considerations~\cite{Csorgo:2016ypf}. 
A possible choice for the  temperature dependent bulk viscosity over entropy density ratio,
that may have a sharp peak at certain value of the critical temperature, as given by eqs.~(1) and~(2) 
of ref.~\cite{Bozek:2017kxo}. Such a particular temperature dependent kinematic bulk viscosity 
corresponds to one of the possible special cases: according to eqs.~\eqref{eq:entropy_equation_kT_mu_eq_0}-\eqref{eq:simplified-kinematic-bulk}, such a form is 
one of the particular cases within our family of exact solutions of viscous relativistic hydrodynamics.

\section{Asymptotically perfect solutions}
\label{sec:7}
\begin{figure}
    \centering
    \resizebox{0.5\textwidth}{!}{%
    \includegraphics{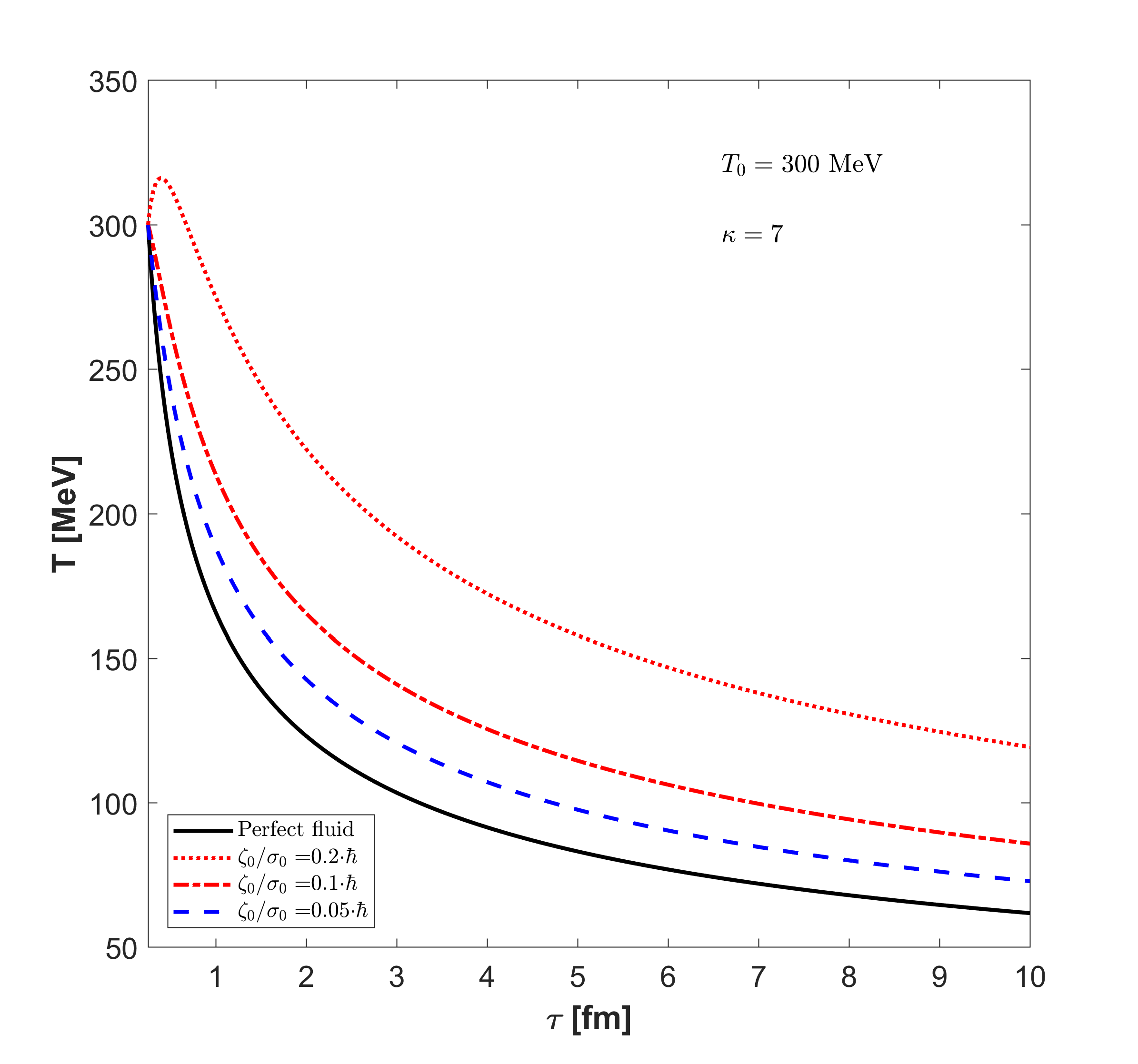}}
    \caption{The evolution of the temperature as a function of the proper time for the solution of Navier-Stokes equation for a vanishing baryochemical potential, $T_0=300$ MeV initial temperature and a constant, temperature independent speed of sound, $c_s^2 = 1/\kappa = 1/7$. The solid black line stands for the CCKH perfect fluid solution, the dashed blue, the dotted-dashed red and the dotted red lines correspond to our new viscous solution of Navier-Stokes equation for different values of kinematic viscosity, but for the same initial conditions.
    }
    \label{fig:temperature_NS_T0fix}
\end{figure}

\begin{figure}
    \centering
    \resizebox{0.5\textwidth}{!}{%
    \includegraphics{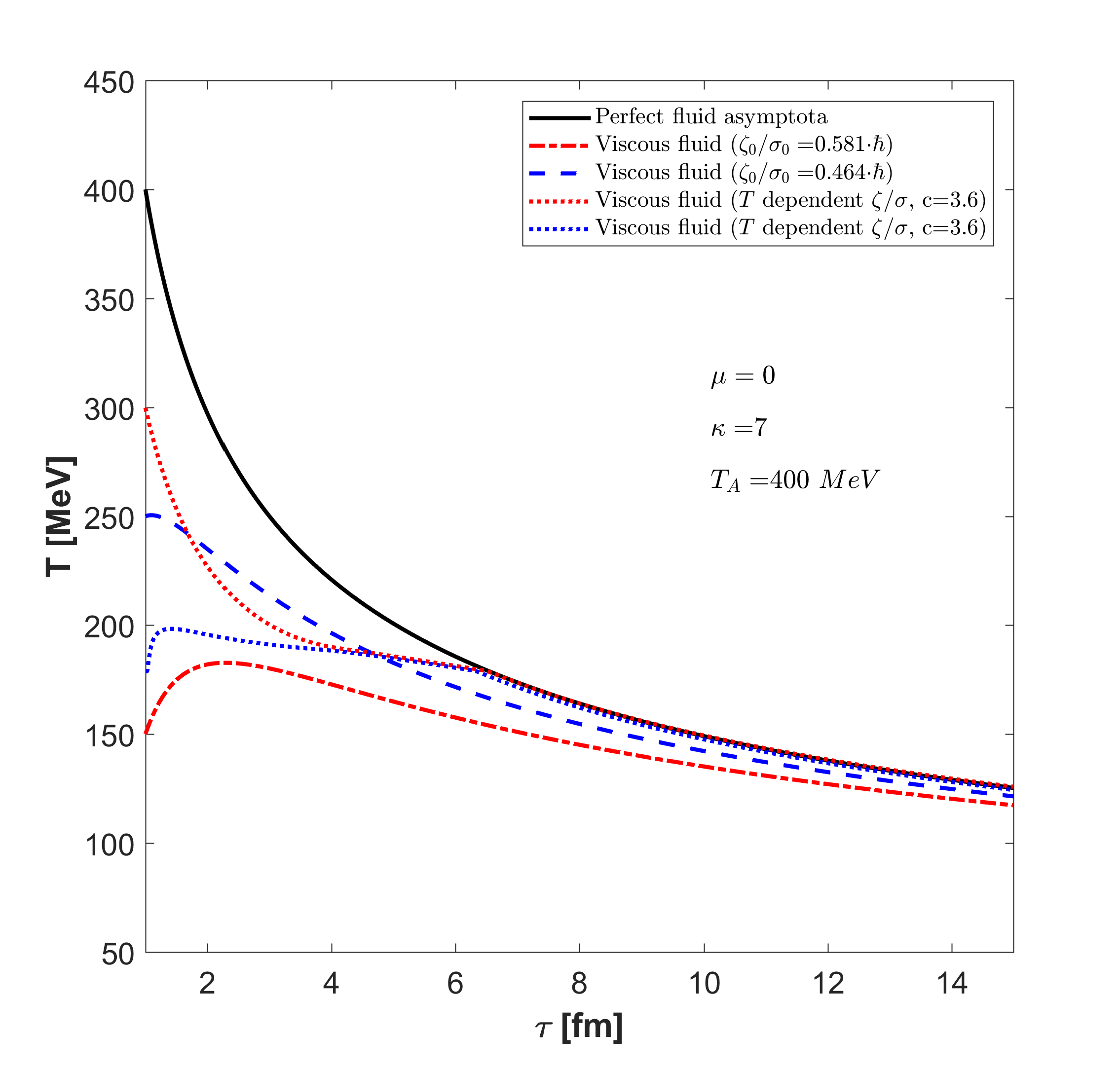}}
    \caption{The evolution of the temperature as a function of the proper time for various viscous solutions of the relativistic Navier-Stokes equations, for a vanishing baryochemical potential and a constant, temperature independent speed of sound, $c_s^2 = 1/\kappa = 1/7$.     
    The solid black line stands for     a perfect fluid solution with $T_A = 400$ MeV. 
    This curve is  approached by each of the shown exact viscous solutions asymptotically, in the $\tau\rightarrow\infty$ limit.
    The blue dashed line shows the time evolution of the temperature for a temperature independendent kinematic bulk viscosity,
    with $T_0 = 250$ MeV and $T_A = 400$ MeV, corresponding to eqs.~(\ref{eq:temp_solution},\ref{eq:temp-solution-zeta0}).
    The dot-dashed red line shows the same but for $T_0 = 150$ MeV.
    The dotted red line shows a calculation with a temperature dependent kinematic bulk
    viscosity, that peaks around the critical temperature, using the parametrization of ref.~\cite{Bozek:2017kxo}, but with a peak that is enhanced by a factor of $c= 3.6$, and with an initial temperature $T_0 = 300$ MeV.
    The dotted blue  line shows the same, but with an initial temperature very close to $T_c$, where the bulk viscosity is expected to have a peak.
    }
    \label{fig:temperature}
\end{figure}

Let us examine the asymptotic behaviour of our new solutions, and with that how the effect of bulk viscosity appears 
at low temperatures. In the $\tau \gg \tau_0$ limit, in both cases of $\mu\neq 0$ and $\mu=0$ 
with $\kappa(T) = \kappa_0$ lead to the same asymptotic perfect fluid temperature profile:
\begin{equation}\label{eq:asymp_T}
    T \sim T_A \left(\frac{\tau_0}{\tau}\right)^{\frac{d}{\kappa_0}}\mathcal{T}\left(s_x,s_y,s_z\right).
\end{equation}
If $\mu = 0$, the entropy density asymptotically equals to a perfect fluid form:
\begin{equation}\label{eq:asymp_sigma_mu0}
    \sigma \sim \sigma_A \left(\frac{\tau_0}{\tau}\right)^{d}\mathcal{V}\left(s_x,s_y,s_z\right).
\end{equation}
At $\mu\neq 0$, we keep eq.~\eqref{eq:part_density} for the particle density. If we start the hydro evolution from the same initial conditions but vary the dissipation and the equations of state, the same initial conditions lead to different asymptotic final states as shown in Fig.~\ref{fig:temperature_NS_T0fix}. This is a conventional feature of our solution.
But the most surprising feature of this solution, that in the $\tau\gg \tau_0$ limit, the effect of bulk viscosity can be absorbed to an asymptotic normalization constant, and we re-obtain the perfect fluid solutions of ref.~\cite{Csorgo:2003rt}. Accordingly, we cannot decide from final state measurements that the medium evolved as a perfect fluid with higher initial temperature ($T_A$) or as a viscous fluid with lower initial temperature ($T_0$). This effect is illustrated in Fig.~\ref{fig:temperature}: in other words, fixing the asymptotic perfect fluid solution and varying the dissipation corresponds to co-varied initial conditions, that lead to the same final state.
As far as we know, the first exact solutions of the relativistic Navier-Stokes hydrodynamics for non-vanishing bulk viscosities were reported in ref.~\cite{Csanad:2019lcl}. Those results support our conclusions.
Three of the five   cases investigated there (cases C, D and E of ref.~\cite{Csanad:2019lcl}),
are also asymptotically  perfect, just like the solutions presented in this work.
One of the five cases (Case B of ref.~\cite{Csanad:2019lcl}) is apparently not physical, while the remaining case (Case A of 
ref.~\cite{Csanad:2019lcl}) is exceptional as it  does not  have an asymptotically perfect form.

On Fig.~\ref{fig:temperature}, we show the asymptotic CCHK perfect fluid solution~\cite{Csorgo:2003ry} with $T_A = 400$ MeV,
together with two different initial temperature for the evolution of a constant kinematic bulk viscosity,  with $T_0 = 250$ and $300$ MeV,
as indicated by the dot-dashed red and the dashed blue lines, respectively, illustrating eqs.~(\ref{eq:temp_solution},\ref{eq:temp-solution-zeta0}) 
that imply that several different initial conditions with different values of $T_0$ may lead to the
same asymptotic perfect fluid behaviour, as the value of $T_0$ and $\zeta_0/\sigma$ both cancel from the asymptotic CCKH solution,
the only important physical characteristics at late time is given by $T_A$ for a given equation of state $\kappa$ and baryochemical potential
$\mu = 0$. The numerical value of $T_0$ was chosen to show a small reheating effect and to indicate that in this case a nearly flat, proper-time independent constant $T\approx T_c$ temperature can be reached, similarly to the case of a nearly equilibrated first order phase transition.
On the same Fig.~\ref{fig:temperature}, we also show the proper-time evolution for a temperature dependent
kinematic bulk viscosity, using the parametrization of ref.~\cite{Bozek:2017kxo}, but with a peak that is enhanced by a factor of $c= 3.6$
to show more clearly the bulk viscosity effects. Such a narrow temperature range where the bulk viscosity is dominant is effectively decreasing the temperature range where reheating effects due to bulk viscosity are important. This can be  clearly and exactly shown analytically: If  the peak
of the kinematic bulk viscosity can be idealized in eq.~(\ref{eq:energy_equation_kT_mu_eq_0}) as 
\begin{equation}
\frac{\zeta}{\sigma}(T) = T_c \frac{\zeta_c}{\sigma_c} \, \delta(T-T_c) ,
\end{equation}
then the temperature will remain constant at $T = T_c$ for a proper-time duration of $\Delta \tau$ that is a monotonic increasing function
of $\frac{\zeta_c}{\sigma_c}$. Such a behaviour is mimicking both qualitatively and quantitatively the Maxwell-Boltzmann construction effect for a first order phase transition. This implies that the physical effects of a large and narrow peak of kinematic bulk viscosity
will be similar to the physical effects expected at a strong first order phase transition. The re-heating and subsequent nearly constant temperature due to a strong peak in bulk viscosity may thus give similar observables as the nucleation scenario in a first order QCD phase transition,
discussed by Csernai and Kapusta in ref.~\cite{Csernai:1992as} and may also lead to a sudden, nearly simultaneous hadron-flash
with signatured discussed in ref.~\cite{Csorgo:1994dd,Csernai:1995zn,Rafelski:2000by}. The detailed description of such an effective first order
phase transition due to bulk viscosity effects, however, goes well beyond the scope of this manuscript.

\section{Exact solution of the Israel-Stewart type theory with Hubble-flow}
\label{sec:8}

\begin{figure}
    \centering
    \resizebox{0.48\textwidth}{!}{%
    \includegraphics{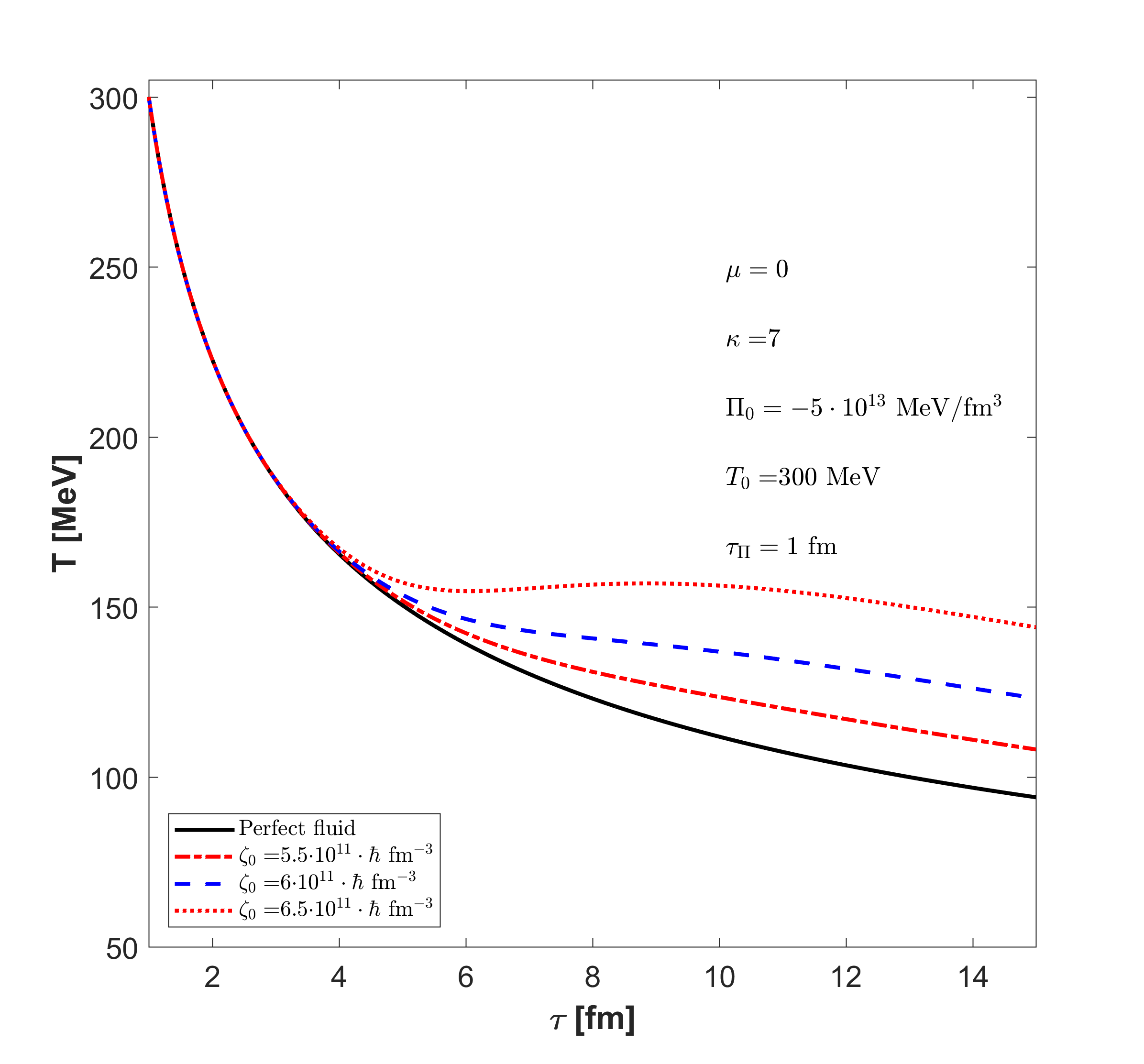}}
    \resizebox{0.48\textwidth}{!}{%
    \includegraphics{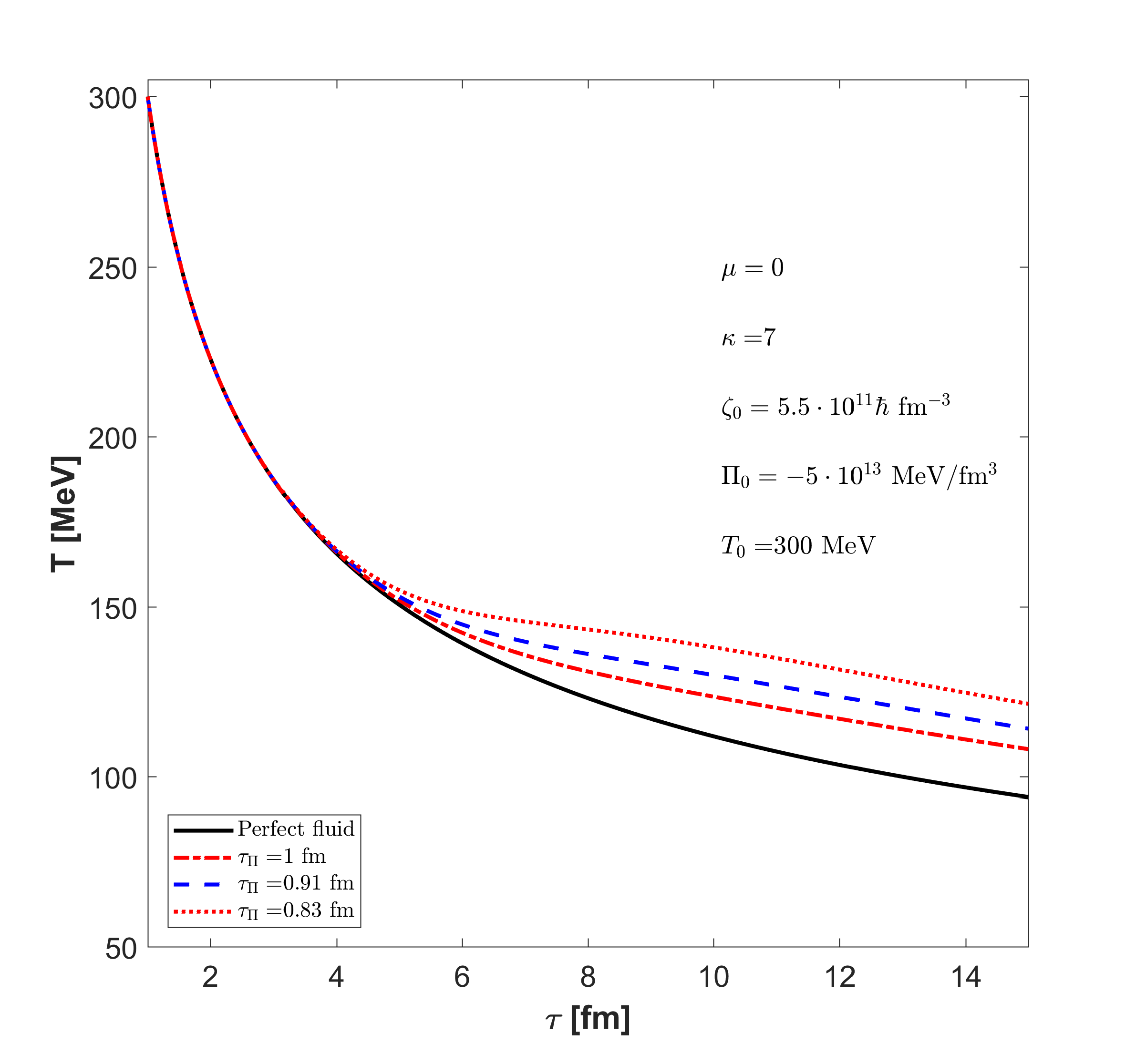}}
    \caption{The evolution of the temperature as a function of the proper time for the solution of Israel-Stewart equation for a vanishing baryochemical potential, $T_0=300$ MeV initial temperature, and a constant, temperature independent speed of sound, $c_s^2 = 1/\kappa = 1/7$. The solid black line stands for the CCKH perfect fluid solution, the dashed blue, the dotted-dashed red and the dotted red lines correspond to our new viscous solution of Israel-Stewart type theory for different values of $\zeta_0$ with a fixed $\tau_{\Pi}$ (upper panel), and for different values of $\tau_{\Pi}$ with a fixed $\zeta_0$ (bottom panel).
    }
    \label{fig:temperature_IS_T0fix}
\end{figure}

\begin{figure}
    \centering
    \resizebox{0.48\textwidth}{!}{%
    \includegraphics{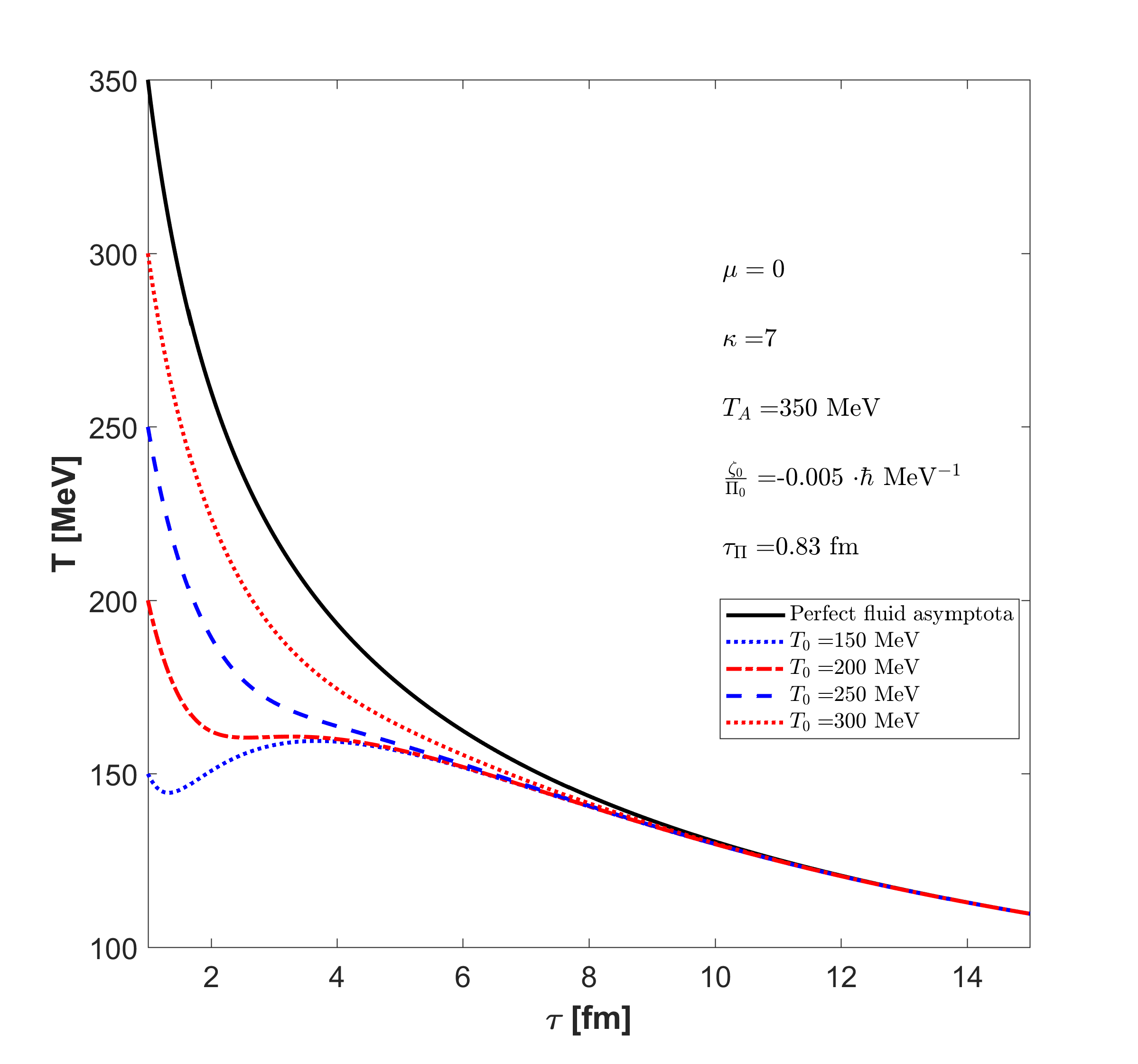}}
    \resizebox{0.48\textwidth}{!}{%
    \includegraphics{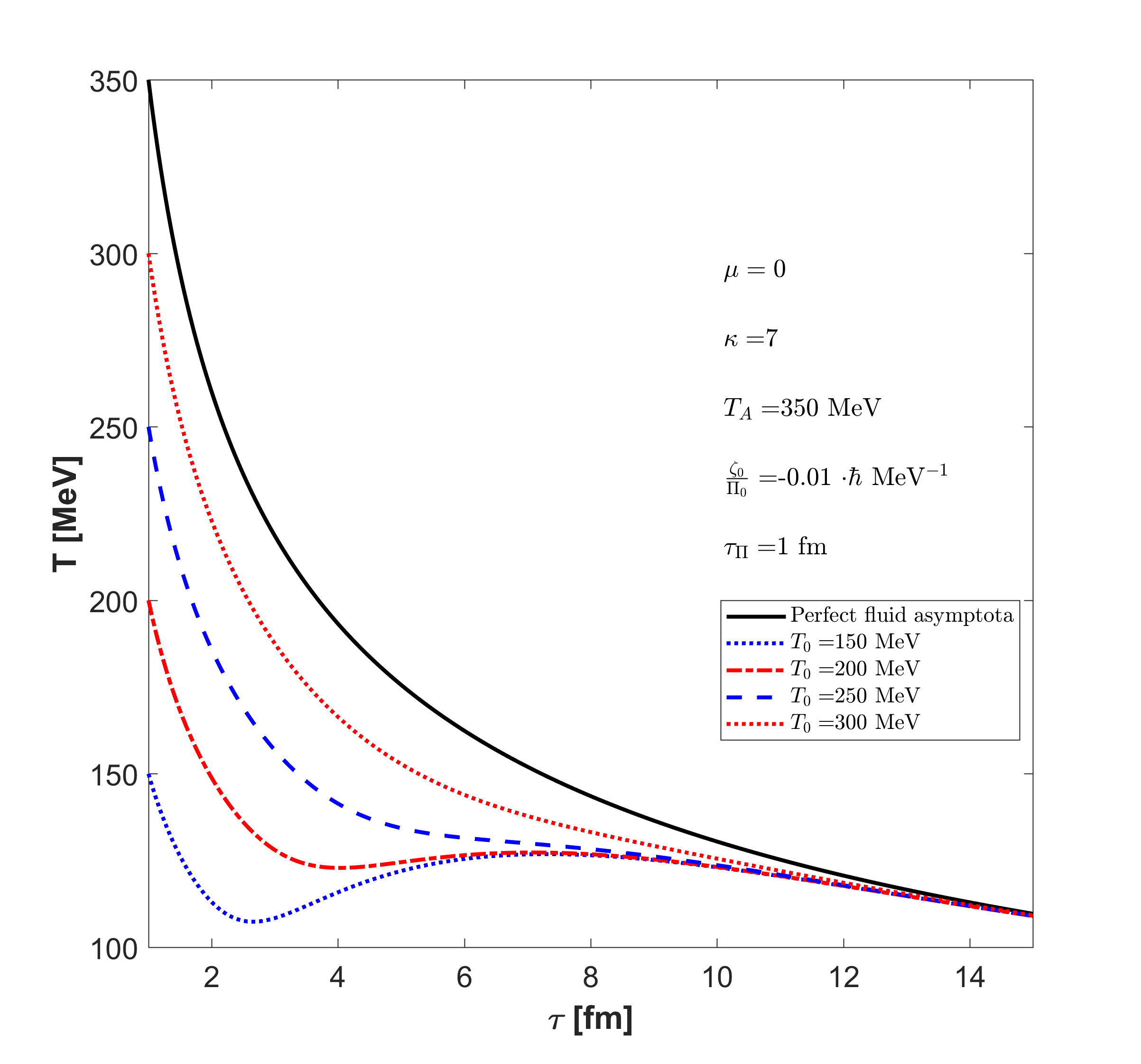}}
    \caption{The evolution of the temperature as a function of the proper time for the solution of Israel-Stewart equation for a vanishing baryochemical potential and a constant, temperature independent speed of sound, $c_s^2 = 1/\kappa = 1/7$. We also fixed the value of $\zeta_0/\Pi_0$ and $\tau_{\Pi}$. In the upper panel we used $\zeta_0/\Pi_0=-0.005\cdot\hbar$ MeV$^{-1}$ and $\tau_{\Pi}=0.83$ fm, while in the bottom panel we set these parameters to $\zeta_0/\Pi_0=-0.01\cdot\hbar$ MeV$^{-1}$ and $\tau_{\Pi}=1$ fm. 
    The solid black line stands for the CCKH perfect fluid solution with $T_A=350$ MeV. It is approached asymptotically in the $\tau\rightarrow\infty$ limit by our new, exact solution of Israel-Stewart hydrodynamics, shown by dotted blue, dotted red, dotted-dashed red and dashed blue lines for different $T_0$ initial conditions.
    }
    \label{fig:temperature_IS_TAfix}
\end{figure}

If the equation of state has the form like $\varepsilon=\kappa p$ with constant $\kappa$, and in the Israel-Stewart type theory~\cite{deSouza:2015ena} we utilize a Hubble flow field, the energy conservation equation reduces to the following 
differential equation~\cite{Csanad:2019lcl}:
\begin{equation}\label{eq:energycons_ingeneral}
    \dot{p}+\frac{\kappa+1}{\kappa}\frac{d}{\tau}p = -\frac{d}{\tau}\frac{\Pi}{\kappa},
\end{equation}
where $\Pi\equiv\Pi(\tau)$ is the bulk pressure, an arbitrary function of $\tau$.  According to ref. ~\cite{Csanad:2019lcl},
the bulk pressure $\Pi  $ is determined by the following first order, ordinary differential equation \cite{deSouza:2015ena}:
\begin{equation}\label{eq:pi_diffeq}
   \Pi = -\zeta\frac{d}{\tau} - \tau_{\Pi} \dot{\Pi},
\end{equation}
$\tau_{\Pi}$ is a relaxation parameter and in the $\tau_{\Pi}\rightarrow 0$ limit the above equation leads back to the relativistic Navier-Stokes approach. In this section, we assume that the bulk viscosity is proportional to the bulk pressure:
\begin{equation}
    \zeta=\Pi \frac{\zeta_0}{\Pi_0},
\end{equation}
where $\Pi_0$ is defined as $\Pi_0=\Pi(\tau=0)$ and its value has to be non-positive to ensure the non-negative entropy production in the entropy balance equation:
\begin{equation}
    \partial_{\tau}\sigma + \frac{d}{\tau}\sigma = -\frac{d}{\tau}\frac{\Pi}{T}
\end{equation}
Using the $\zeta \propto \Pi$ ansatz eq.~\eqref{eq:pi_diffeq} becomes very simple to solve:
\begin{equation}
    \Pi(\tau)=\Pi_0\left(\frac{\tau_0}{\tau}\right)^{\frac{d}{\tau_{\Pi}}\frac{\zeta_0}{\Pi_0}}\exp\left(-\frac{\tau-\tau_0}{\tau_{\Pi}}\right).
\end{equation}
If $\Pi(\tau)$ is embedded into eq.~\eqref{eq:energycons_ingeneral}, then we obtain a first order, inhomogeneous, ordinary differential equation for $p$, and its solution reads as:
\begin{equation}
    p(\tau)=p_A \left(\frac{\tau_0}{\tau}\right)^{d\left(1+\frac{1}{\kappa}\right)}\left[1+\frac{p_0-p_A}{p_A}\cdot\frac{\Gamma\left(B,\frac{\tau}{\tau_{\Pi}}\right)}{\Gamma\left(B,\frac{\tau_0}{\tau_{\Pi}}\right)} \right],
\end{equation}
where $\Gamma$ stands for the incomplete gamma-function, we denoted the initial pressure with $p_0\equiv p(\tau=0)$, $p_A$ and $B$ are the following constants:
\begin{align}
p_A&=p_0-\frac{\Pi_0 d}{\kappa}\left(\frac{\tau_0}{\tau_{\Pi}}\right)^{-B}\exp\left(\frac{\tau_0}{\tau_{\Pi}}\right)\Gamma\left(B,\frac{\tau_0}{\tau_{\Pi}}\right),
\\
B&=d\left(1+\frac{1}{\kappa}-\frac{\zeta_0}{\Pi_0}\frac{1}{\tau_{\Pi}}\right).
\end{align}
The details of the derivation of this solution are presented in ref.~\cite{Csanad:2019lcl}. Here we stress that the above exact solution of the IS theory
also belongs to the class of asymptotically perfect solutions of relativistic dissipative hydrodynamics.
Here $p_A$ has the same physical meaning as before: it is the initial pressure of an asymptotic perfect fluid solution, found by Csörgő, Csernai, Kodama and Hama \cite{Csorgo:2003ry}. Keep in mind, that $p_A > p_0$, because $\Pi_0 \le 0$.
In Fig.~\ref{fig:temperature_IS_T0fix} we show if the initial conditions are fixed, and the bulk viscosities and relaxation parameters are varied, then different hadronic final states are emerged. In addition, we have found that for late times ($\tau\gg \tau_0$) not only the solution of the Navier-Stokes equation, but also the solution of the Israel-Stewart equation approaches the CCKH solution with $p_A$ initial pressure:
\begin{equation}\label{eq:IS_limes}
    \lim\limits_{\tau\rightarrow\infty} p(\tau) = p_A\left(\frac{\tau_0}{\tau}\right)^{d\left(1+\frac{1}{\kappa}\right)}.
\end{equation}
In conclusion the Israel-Stewart theory also has exact solutions in the class of asymptotically perfect exact solutions, where at late times the effects of dissipation enter through a scaling variable ($p_A$) only.
We illustrate this behaviour in Fig.~\ref{fig:temperature_IS_TAfix}: fixing the asymptotic perfect fluid solution and varying the dissipative coefficient corresponds to co-varied initial conditions. If the initial conditions and the dissipative coefficients are properly co-varied, then the same asymptotic final state can be obtained. 

\section{Discussion}

In this paper we deal with a class of exact solutions of dissipative relativistic hydrodynamics, referred to as the class of asymptotically perfect exact solutions. The simplest form of these solutions is presented in Section~\ref{sec:3}.

We are interested in following our academic interest and thus we have investigated in the previous sections, how broad the class of asymptotically perfect solutions of dissipative relativistic hydrodynamics may be. In Section \ref{sec:4}. we have shown that this class includes solutions with fluctuating initial density profiles for the case when a conserved charge exist. In Section~\ref{sec:5} we have  shown that fluctuating initial temperature profiles are also included in this class even in the case when conserved charges do not play a role, such as in the early universe or in ultra-relativistic heavy ion collisions with approximately vanishing baryochemical potentials. In Section ~\ref{sec:6} we have shown,
that the class of asymptotically perfect solutions includes also exact solutions with a temperature dependent speed of sound. The asymptotically perfect type of these solutions is detailed in Section~\ref{sec:7}. However, the analysis of the non-relativistic limit of these solutions
as well as the comparison of these solutions with the experimental data is well beyond the scope of the present manuscript.
%, it will be a topic of a follow-up paper.

This class of asymptotically perfect solutions is apparently rather broad as it includes exact solutions both to the Israel-Stewart and the Navier-Stokes theory, and as we have shown
in Section~\ref{sec:8}. As this class also includes solutions with fluctuating initial temperature profiles, temperature dependent speed of sound and temperature dependent bulk and shear viscosity coefficients, actually this class may have a great academic interest.

We have shown that it is possible to co-vary the initial conditions and the equations of state  with viscosity and dissipative coefficients so that the asymptotic perfect fluid solution remains the same: in the $\tau\rightarrow \infty$ limit the temperature, the entropy density and the pressure of our solution approaches that of the CCKH perfect fluid solution. However, if the initial conditions of the perfect fluid and the viscous solutions are the fixed (the same), then the perfect fluid solutions of course deviate  from the solutions of dissipative relativistic hydrodynamics and one may think that the hadronic final states can carry information about the dissipative coefficients, because of the entropy production. In Fig.~\ref{fig:temperature_NS_T0fix} and in Fig.~\ref{fig:temperature_IS_T0fix} we demonstrated this - rather expected and standard - behaviour. 

People solving the equations of dissipative relativistic hydrodynamics typically try to fit experimental data and extract the transport coefficients like shear and bulk viscosity or $\tau_{\Pi}$ of Israel-Stewart theory from these data, under the assumption that their solutions are not in the class of asymptotically perfect solutions.  This is the specific wide-spread belief, however, by numerical investigations it is very difficult to 
investigate if a solution is asymptotically perfect or not, as the numerical viscosity makes the investigation of the asymptotic behaviour rather difficult. Although we have demonstrated exactly that this class of asymptotically perfect solutions exists and is rather broad, without detailed investigations we cannot tell a priorily if the experimentally relevant solutions belong to this class or not. However, if an (academically interesting) solution is in this class, then due to eqs.~\eqref{eq:IS_limes} for the Israel-Stewart theory and also $p_A$ and $\sigma_A$ of eqs.~\eqref{eq:pressure_scale} and \eqref{eq:entropy_scale} only a combination of the initial conditions and the dissipative coefficients can be determined from the late time behaviour of these exact solutions. The result is a new kind of final state scaling of dissipative relativistic hydrodynamics. The possibility to co-vary the initial conditions with the dissipative coefficients to reach the same final state in
the case of relativistic NS theory is illustrated on Fig. ~\ref{fig:temperature} while the equivalent behaviour of the IS theory is illustrated on ~\ref{fig:temperature_IS_TAfix}.

In this manuscript we have provided a hopefully rather clear mathematical result about the existence of a new class of asymptotically perfect solutions of dissipative relativistic hydrodynamics. Previously the existence of such a class was not even imagined. In Section \ref{sec:7}
we have shown that in the $\tau \gg \tau_0$ limit, the effect of bulk viscosity can be absorbed to an asymptotic normalization constant, so the CCKH perfect fluid solution is re-obtained. We do not claim, however, that experimental data corresponds to the behaviour of the exact solutions.
For the time being the existence of the asymptotically perfect class of solutions is an interesting but rather academic result. The comparison with experimental data goes well beyond the scope of the present, already rather extended manuscipt.

We hope that this discussion clarified the main result of this paper sufficiently well. By including the solutions of the Israel-Stewart theory with the same property of asymptotic perfectness, we faithfully demonstrated that there exists a broad class of asymptotically perfect solutions of dissipative relativistic hydrodynamics where the asymptotic, late time behaviour of the fluid behaviour is dependent only on a scaling variable $p_A$ (or equivalently $T_A$ or $\sigma_A$), the pressure (or, temperature or entropy density) of an equivalent perfect fluid solution. The combination $p_A$ (or $T_A$ or $\sigma_A$) determines the final state observables in this class of solutions. Different initial conditions, different equations of state and different dissipative coefficients yield exactly the same asymptotic solutions, if the combinations $p_A$ (or $T_A$ or $\sigma_A$) are exactly the same.

\section{Summary}
\label{sec:9}
We have found a new family of analytic, exact solutions of relativistic Navier-Stokes and also Israel-Stewart hydrodynamics in  1+3 dimensions. These solutions are causal and asymptotically perfect. These solutions feature a Hubble-type velocity profile, both for a finite and for a vanishing baryochemical potential as well.
%under the conditions that the bulk viscosity to pressure ratio is a constant, and the speed of sound is a function of the temperature only. 
The effects of heat conduction $\lambda(T)$ and shear viscosity $\eta(T)$ cancel from these solutions. %, exactly.
Thus our new solutions can describe the medium for any value of 
$\lambda(T)$, 
$\eta(T)$ and bulk viscosity $\zeta(T)$. This is useful to test numerical algorithms and solutions. 

The asymptotic effects of bulk viscosity were also examined here. We have found that in the $\tau\gg \tau_0 $ limit the 
entropy production becomes negligibly small, and the time evolution approaches that of a perfect fluid. 
We find that this  feature is characteristic of several of the solutions of ref.~\cite{Csanad:2019lcl}, too.
Thus
no mathematically precise information can be gained about viscosity effects by studying only
the hadronic final state of the asymptotically perfect fluids in relativistic heavy ion collisions,
if the velocity field has a special, asymptotically Hubble form.

We find that the shear viscosity and  heat conductivity cancels from these solutions exactly,
while the bulk viscosity cancels asymptotically. Thus,  for sufficiently late times,
our new solutions asymptotically approach the exact solutions of perfect fluid hydrodynamics.
In that sense, our solutions provide a mathematically exact counter-example for several wide-spread beliefs
supported by detailed numerical analysis of comparisons of numerical solutions of relativistic viscous hydrodynamics and
experimental data~\cite{Sangaline:2015isa,Pratt:2015zsa}, but lacking the support of mathematically exact results that are presented here.
The entropy production during the time evolution of the viscous fluid in our case can thus be absorbed to a higher initial temperature for the  time evolution of an asymptotically equivalent perfect fluid. Asymptotically, exactly the same hadronic final state can be reached from both the viscous and the perfect evolution scenarios, even for multipole solutions with fluctuating initial temperature and entropy density profiles.
This strengthens earlier but not so well-known exact results, 
presented typically for directional Hubble flows, that indicated that co-varying the equation of state and the initial
temperature may result in exactly the same hadronic final state~\cite{Csanad:2005gv,Csanad:2007iv,Nagy:2016uiz},
similarly to the exact results that were obtained also for rotating exact solutions of perfect fluid fireball hydrodynamics~\cite{Nagy:2016uiz,Csorgo:2018tsu}.
One of our surprising analytic results is  that the physical effects of a large and narrow peak of kinematic bulk viscosity
are found to be very similar to the effects expected at a strong first order phase transition,
as discussed in refs.~\cite{Csorgo:1994dd,Csernai:1995zn,Rafelski:2000by}
%as in this case the temperature
%remains nearly constant within a narrow range of the critical temperature $T_c$, as illustrated by the dotted blue line 
and illustrated on Fig.~\ref{fig:temperature}.

Applications of these Hubble-type non-perfect fluid solutions may be expected in
the field of relativistic magnetohydrodynamics, based on the recent great progress of solutions in 1+1 dimensional magnetohydrodynamical problems~\cite{Shokri:2018qcu,Sadooghi:2018fkg,She:2019wdt,Moghaddam:2020dix}.
Further, detailed applications of our exact theoretical 
results to the analysis of experimental data are possible, but go well beyond the
scope of our current manuscript. 

%The non-relativistic kinematic limit of our new solution is also of certain interest as apparently new class of exact solutions
%of the non-relativistic Navier-Stokes problem can be obtained by taking the non-relativistic limit of our results.
%In this work, we thus have improved  the level of  our theoretical understanding of some rather difficult parts of the Navier-Stokes Millennium problem of the Clay Mathematical Institute.

\section*{Acknowledgments} We would like to thank particularly to M. Csan\'ad, as well as to Z.-F. Jiang, S. L\"ok\"os and M. I. Nagy for inspiring discussions.
This reasearch was partially supported by the NKFIA grants K 133046 FK 123842 and FK 123959,
the EFOP 3.6.1-16-2016-00001 grant (Hungary), and THOR, the EU COST
Action CA15213.

\clearpage
\bibliographystyle{unsrt}
\bibliography{Rel-Navier-Stokes}

%\begin{table}
%\begin{tabular}{lll}
%\hline\noalign{\smallskip}
%\end{tabular}
% Or use
%\vspace*{5cm}  % with the correct table height
%\end{table}
%
% BibTeX users please use
% \bibliographystyle{}
% \bibliography{}
%\begin{thebibliography}{}
%Author, \textit{Book title} (Publisher, place year) page numbers
% etc
%\end{thebibliography}

\end{document}